# Dielectric-Modulated Impact-Ionization MOS (DIMOS) Transistor as a Label-free Biosensor

N. Kannan, *Member, IEEE*, and M. Jagadesh Kumar, *Senior Member, IEEE*

*Abstract*—In this letter, we propose a dielectric-modulated Impact-Ionization MOS (DIMOS) transistor based sensor for application in label-free detection of biomolecules. Numerous reports exist on the experimental demonstration of nanogap-embedded FET-based biosensors, but an I-MOS based biosensor has not been reported previously. The concept of a dielectric-modulated I-MOS based biosensor is presented in this letter based on TCAD simulation study. The results indicate a high sensitivity to the presence of biomolecules even at small channel lengths. In addition, a low variability of the sensitivity to the charges on the biomolecule is observed. The high sensitivity, dominance of dielectric-modulation effects and operation at even small channel lengths makes the DIMOS biosensor a promising alternative for CMOS-based sensor applications.

## I. INTRODUCTION

The application of FET as a biosensor for label-free detection of charged biomolecules has seen significant usage [1-3]. The use of the concept of dielectric modulation of a vertical nanogap in the FET's gate due to the presence of biomolecules has enabled the application of FET biosensors for detecting the presence of charge-free biomolecules as well [4-8]. The reported dielectric-modulated FET (DMFET) based biosensors show high responsiveness to both dielectric modulation and charge of the biomolecules, with the two effects often affecting the device parameters in opposite directions leading to reduced sensitivity [8]. In addition, the DMFET shows a sharp dependence of the sensitivity of biomolecule detection to the nanogap length and a significant fall in sensitivity for short channel lengths [8]. In this letter, we propose, for the first time, the application of impact-ionization MOS (I-MOS) transistor as a bio-sensor. A conventional I-MOS is a gated p-i-n diode, with the gate voltage modulating the channel inversion charge and causing the drain-source voltage to appear across the intrinsic region outside the gate region [9, 10]. On reaching a critical value, the electrical field in the intrinsic region causes impact-ionization leading to an abrupt turning-ON of the I-MOS transistor [11, 12]. Recent results have shown a sub-threshold slope as low as 2-10 mV/dec [13-15].

This work was supported in part by NXP (Philips) Chair Professorship awarded to M.J. Kumar. The authors are from the Department of Electrical Engineering, Indian Institute of Technology, Delhi, Hauz Khas, New Delhi 110016, India.
Email: n.kannan@outlook.com; mamidala@ee.iitd.ac.in

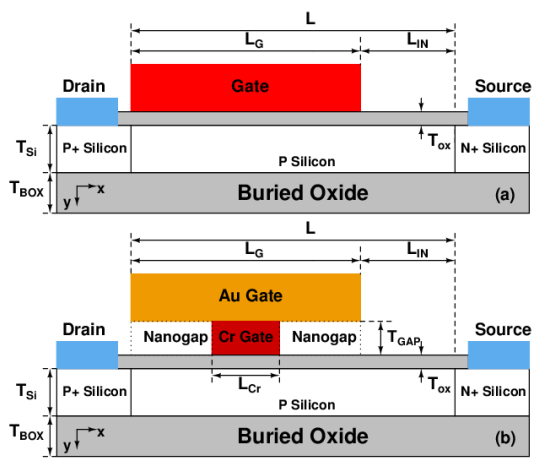

Fig. 1. Cross sectional view of (a) the conventional I-MOS and (b) the DIMOS.

In this letter, using TCAD simulation [16], we examine the performance of a Dielectric modulated I-MOS transistor (DIMOS) as a label-free biosensor. By introducing a nanogap using a combination of gold and chromium gate structure [4] in the I-MOS device, we demonstrate the DIMOS to be a highly sensitive bio-sensor with potential applications in bio-medical field for sensing bio-molecules.

## II. DEVICE STRUCTURE AND BIOSENSING ACTION

The structure of the conventional p-channel I-MOS [9, 10] and p-channel DIMOS are shown in Fig. 1(a) and Fig. 1(b), respectively. The parameters used in our simulation are: silicon film thickness ($T_{Si}$) = 100 nm, silicon film doping = $2 \times 10^{15}$ cm$^{-3}$, buried oxide thickness ($T_{BOX}$) = 300 nm, gate-oxide thickness ($T_{ox}$) = 5 nm, the nominal channel length (L) = 200 nm, nanogap thickness ($T_{GAP}$) = 15 nm, chromium gate length ($L_{Cr}$) = 25 nm, work function of chromium (Cr) = 4.5 eV, gold gate length ($L_G$) = 125 nm and work function of gold (Au) = 5.1 eV. For the conventional I-MOS, the additional parameters used are: the gate length ($L_G$) = 125 nm and the gate work function ($\Phi_M$) = 4.6 eV. For both the devices, the length of the intrinsic region ($L_{IN}$) outside the gate region is 75 nm. The doping profiles in the source and drain region are modeled as a Gaussian distribution with the drain acceptor doping ($N_A$) and source donor doping ($N_D$) taken as $10^{20}$ cm$^{-3}$.

The creation of the nanogap in the gate structure has been demonstrated experimentally in [4]. After creating the nanogap by etching the edges of the chromium gate structure, air, with a dielectric constant (K) of 1, fills the nanogap [8]. Immobilization of biomolecules (K > 1) in the nanogap causes the effective dielectric constant to increase, leading to an increased gate capacitance and consequently a reduced threshold voltage $V_T$. Detection of this $V_T$ shift leads to a label-free detection of biomolecules [4]-[8].

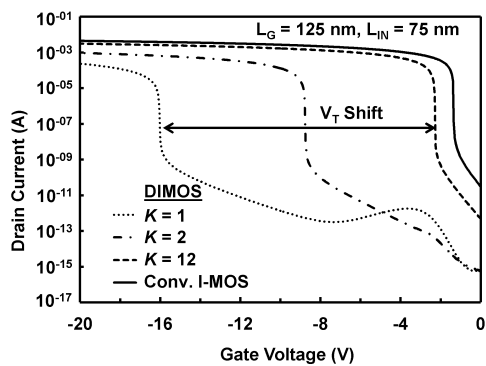

Fig. 2. P-channel DIMOS $I_D$ vs. $V_G$ characteristics.

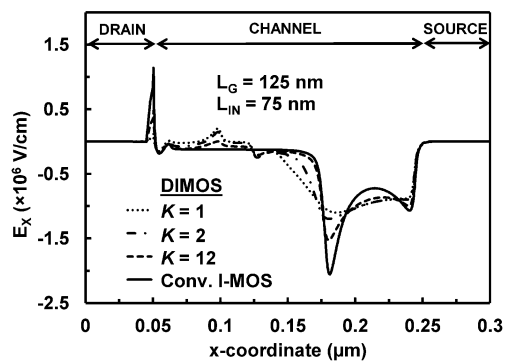

Fig. 3. Horizontal electric field ($E_X$) profile along the x-axis.

## III. SIMULATION RESULTS AND DISCUSSION

We have simulated the DIMOS biosensor to study the sensitivity of the device for detection of the immobilization of the nanogap by biomolecules. The immobilization event is modeled in the simulation by changing the dielectric constant of the nanogap since the biomolecules will have a dielectric constant different from air. To represent a group of biomolecules with low and high dielectric constants, a value of $K = 2$ and 12 is used, respectively, following the approach in [8]. We selected the Selberherr's Model for impact-ionization and also included Band-to-Band tunneling effects in the simulation.

Fig. 2. shows the $I_D$ vs. $V_G$ characteristics of the DIMOS biosensor, with $V_S = 7.8$ V and $V_D = 0$ V. A negative sweep is applied on the gate as simulation was done on a p-Channel DIMOS. In the absence of the biomolecular immobilization ($K=1$), the DIMOS turns on at ~16V as compared to the conventional I-MOS turning on at ~1.36V. The higher turn on voltage for the DIMOS, compared to the conventional I-MOS, is because of the increased dielectric thickness in the nanogap region (with K = 1) of the DIMOS. This leads to a reduced gate control of the channel increasing the turn on voltage. The presence of high dielectric constant biomolecules ($K = 12$) causes the $V_T$ of DIMOS to fall to 2.29V (~86% $V_T$ shift), leading to an extremely sensitive biosensor even at small channel lengths. Another parameter to be considered for the device sensitivity is the On-state current $I_{ON}$ which shows an increase by thirteen times in the presence of biomolecules when compared with the empty nanogap.

This large shift in $V_T$ with nanogap immunization can be understood by analyzing the transverse electric field profile along the channel of the DIMOS, as shown in Fig. 3. The I-MOS turns-on because of the impact ionization caused by high horizontal electric field in the intrinsic region [10]. The empty nanogap results in a low inversion in the channel region under the nanogap, causing $V_{SD}$ to appear across a longer length of the silicon film. This results in the impact ionization setting in at a much higher gate voltage. With the biomolecule immunization and consequent dielectric-modulation of the nanogap, sufficient inversion occurs in the channel region under the nanogaps. As a result, $V_{SD}$ now appears across a shorter length of the silicon film. This will make the electric field profile of the DIMOS similar to that of the conventional I-MOS leading to a reduction in $V_T$ of the DIMOS.

### A. Effect of Gate Length ($L_G$) at a fixed Cr Gate Length ($L_{Cr}$)

We simulated the DIMOS biosensor to find the optimum value of $L_G$ such that the sensitivity to biomolecular immobilization is maximized. The variation of $\Delta V_T$ and $\Delta I_{ON}$ for different values of $L_G$ is shown in Fig. 4. For $L_G = 115$ nm, the device seems to be most sensitive if $V_T$ shift is chosen as the sensing parameter. DMFETs with a channel length of 300 nm have been shown to have less sensitivity [8]. However, the proposed DIMOS with a channel length of 200 nm exhibits high sensitivity indicating that this device can be scaled down to smaller dimensions. We have also simulated the DIMOS at a channel length (L) of 1 μm and the biosensing action remains valid, thus showing the applicability of the DIMOS concept across process technology nodes.

### B. Impact of Biomolecular charges

The DMFET biosensing action is significantly affected by the presence of charges in the biomolecules. Depending on the polarity of charges, the device type (PMOS or NMOS) may have to be changed [7]. We studied the impact of biomolecule charges on the performance of the DIMOS biosensors, choosing $L_G = 115$ nm since this is the optimum value for maximum device sensitivity. Fig. 5 shows the variation in $\Delta V_T$ and $\Delta I_{ON}$ for a range of biomolecular charges [5]. The biomolecular charges have been modeled as point charges at the silicon-oxide interface. Compared to the conventional

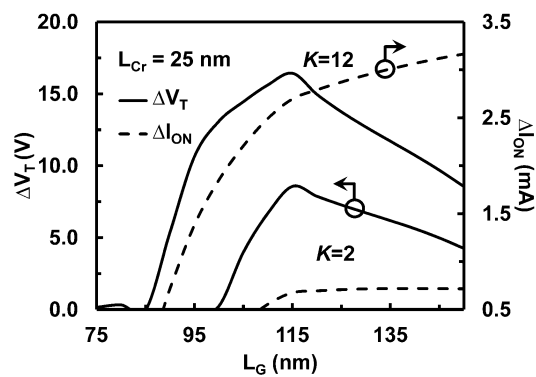

Fig. 4. $\Delta V_T$ and $\Delta I_{ON}$ for the DIMOS as a function of gate length, $L_G$.

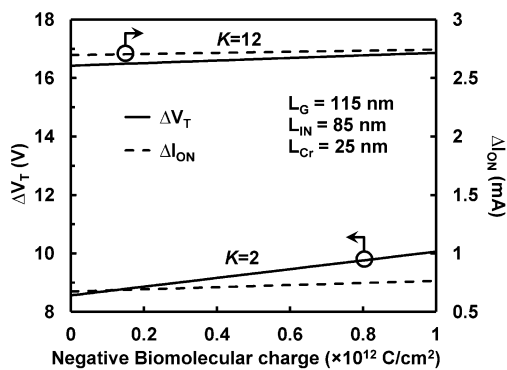

Fig. 5. $\Delta V_T$ and $\Delta I_{ON}$ of the p-channel DIMOS for negatively charged biomolecules.

DMFET where the charge effect can even dominate over the dielectric-modulation effect [7], the DIMOS shows negligible variation in the sensitivity to biomolecular charges. The dominance of the dielectric-modulation effect in DIMOS enables the use of the same device type (p-channel or n-channel) for detecting biomolecules of opposite charge polarities which is not possible in the conventional DMFETs.

## IV. CONCLUSION

We have proposed and analyzed the concept of a dielectric-modulated impact-ionization MOS (DIMOS) biosensor. The proposed structure demonstrates high sensitivity for biomolecule detection and retains dielectric-modulation as the dominant effect, as opposed to the conventional DMFETs where the impact of biomolecule charges on the device sensitivity is significant. The proposed sensor device shows good scalability across process technology nodes and can be an exciting alternative to FET biosensors for both advanced technology nodes as well as the older technology nodes where low-cost is the driving factor.